\documentstyle[emulateapj,psfig,natbib,amsfonts,amsmath,graphicx]{article}
\def\ra#1#2#3{#1$^{\rm h}$#2$^{\rm m}$#3$^{\rm s}$}
\def\dec#1#2#3{$#1^\circ#2'#3''$}
\def\swift{{\it Swift}}
\def\nod{\nodata}

\def\ociw{1}
\def\prince{2}
\def\hubble{3}
\def\psu{4}
\def\uh{5}
\def\cit{6}
\def\la{7}
\def\chi{8}
\def\anu{9}
\def\srl{10}
\def\nrao{11}
\def\pom{12}
\def\oxf{13}
\def\gem{14}
\def\mcgill{15}
\def\car{16}

\begin{document}

\title{A New Population of High Redshift Short-Duration Gamma-Ray
Bursts}

\author{
E.~Berger\altaffilmark{\ociw,}\altaffilmark{\prince,}\altaffilmark{\hubble},
D.~B.~Fox\altaffilmark{\psu},
P.~A.~Price\altaffilmark{\uh},
E.~Nakar\altaffilmark{\cit},
A.~Gal-Yam\altaffilmark{\cit,}\altaffilmark{\hubble},
D.~E.~Holz\altaffilmark{\la,}\altaffilmark{\chi,}\altaffilmark{\ociw},
B.~P.~Schmidt\altaffilmark{\anu},
A.~Cucchiara\altaffilmark{\psu},
S.~B.~Cenko\altaffilmark{\srl},
S.~R.~Kulkarni\altaffilmark{\cit},
A.~M.~Soderberg\altaffilmark{\cit},
D.~A.~Frail\altaffilmark{\nrao},
B.~E.~Penprase\altaffilmark{\pom}
A.~Rau\altaffilmark{\cit},
E.~Ofek\altaffilmark{\cit},
S.~J.~Bell Burnell\altaffilmark{\oxf},
P.~B.~Cameron\altaffilmark{\cit},
L.~L.~Cowie\altaffilmark{\uh},
M.~A.~Dopita\altaffilmark{\anu},
I.~Hook\altaffilmark{\oxf},
B.~A.~Peterson\altaffilmark{\anu},
Ph.~Podsiadlowski\altaffilmark{\oxf},
K.~C.~Roth\altaffilmark{\gem},
R.~E.~Rutledge\altaffilmark{\mcgill},
S.~S.~Sheppard\altaffilmark{\car},
and A.~Songaila\altaffilmark{\uh}
}

\altaffiltext{\ociw}{Observatories of the Carnegie Institution
of Washington, 813 Santa Barbara Street, Pasadena, CA 91101}

\altaffiltext{\prince}{Princeton University Observatory,
Peyton Hall, Ivy Lane, Princeton, NJ 08544}

\altaffiltext{\hubble}{Hubble Fellow}

\altaffiltext{\psu}{Department of Astronomy and Astrophysics,
Pennsylvania State University, 525 Davey Laboratory, University
Park, PA 16802}

\altaffiltext{\uh}{Institute for Astronomy, University of Hawaii,
2680 Woodlawn Drive, Honolulu, HI 96822}

\altaffiltext{\cit}{Division of Physics, Mathematics and Astronomy,
105-24, California Institute of Technology, Pasadena, CA 91125}

\altaffiltext{\la}{Theoretical Division, Los Alamos National
Laboratory, Los Alamos, NM 87545} 

\altaffiltext{\chi}{Kavli Institute for Cosmological Physics and
Department of Astronomy and Astrophysics, University of Chicago,
Chicago, IL 60637}

\altaffiltext{\anu}{Research School of Astronomy and Astrophysics,
ANU, Mt Stromlo Observatory, via Cotter Rd, Weston Creek, ACT 2611,
Australia}

\altaffiltext{\srl}{Space Radiation Laboratory, MS 220-47, California
Institute of Technology, Pasadena, CA 91125}

\altaffiltext{\nrao}{National Radio Astronomy Observatory, Socorro,
NM 87801}

\altaffiltext{\pom}{Pomona College Department of Physics and Astronomy,
610 N. College Avenue, Claremont, CA}

\altaffiltext{\oxf}{Department of Astrophysics, University of Oxford,
Oxford OX1 3RH, UK}

\altaffiltext{\gem}{Gemini Observatory, 670 N. Aohoku Place Hilo, HI
96720}

\altaffiltext{\mcgill}{Department of Physics, McGill University,
Rutherford Physics Building, 3600 University Street, Montreal,
QC H3A 2T8, Canada}

\altaffiltext{\car}{Department of Terrestrial Magnetism, Carnegie
Institution of Washington, 5241 Broad Branch Road NW, Washington, DC
20015}

\begin{abstract} 
The redshift distribution of the short-duration GRBs is a crucial, but
currently fragmentary, clue to the nature of their progenitors.  Here
we present optical observations of nine short GRBs obtained with
Gemini, Magellan, and the {\it Hubble Space Telescope}.  We detect the
afterglows and host galaxies of two short bursts, and host galaxies
for two additional bursts with known optical afterglow positions, and
five with X-ray positions ($\lesssim 6''$ radius).  In eight of the
nine cases we find that the most probable host galaxies are faint,
$R\approx 23-26.5$ mag, and are therefore starkly different from the
first few short GRB hosts with $R\approx 17-22$ mag and $z\lesssim
0.5$.  Indeed, we measure spectroscopic redshifts of $z\approx
0.4-1.1$ for the four brightest hosts.  A comparison to large field
galaxy samples, as well as the hosts of long GRBs and previous short
GRBs, indicates that the fainter hosts likely reside at $z\gtrsim 1$.
Our most conservative limit is that at least half of the five hosts
without a known redshift reside at $z>0.7$ ($97\%$ confidence level),
suggesting that about $1/3-2/3$ of all short GRBs originate at higher
redshifts than previously determined.  This has two important
implications: (i) We constrain the acceptable age distributions to a
wide lognormal ($\sigma\gtrsim 1$) with $\tau_*\sim 4-8$ Gyr, or to a
power law, $P(\tau)\propto\tau^n$, with $-1\lesssim n\lesssim 0$; and
(ii) the inferred isotropic energies, $E_{\gamma,{\rm iso}}\sim
10^{50}-10^{52}$ erg, are significantly larger than $\sim
10^{48}-10^{49}$ erg for the low redshift short GRBs, indicating a
large spread in energy release or jet opening angles.  Finally, we
re-iterate the importance of short GRBs as potential gravitational
wave sources and find a conservative Advanced LIGO detection rate of
$\sim 2-6$ yr$^{-1}$.
\end{abstract}


\section{Introduction}
\label{sec:intro}

The redshift distribution of the short-duration gamma-ray bursts
(GRBs) serves as one of the primary clues to the nature of their
progenitors.  This is because distance measurements determine the
energy budget and its dispersion, provide information on the
progenitor age distribution and its relation to star formation, and
allow us to estimate event rates for gravitational wave detectors such
as LIGO (in the context of NS-NS and NS-BH progenitors).  Initial
observations suggested that short GRBs occur at significantly lower
redshifts than long GRBs (for which $\langle z\rangle\sim 3$; e.g.,
\citealt{bkf+05,jlf+06}).  In particular, GRBs 050724 and likely
050509b are associated with bright ($L\sim 2-4\,L^*$) elliptical
galaxies at $z=0.257$ and $0.226$, respectively
\citep{bpc+05,gso+05,bpp+06,pbc+06}, while GRBs 050709 and 051221a
were localized to star-forming galaxies at $z=0.1606$ and $0.5465$,
respectively, with $L\sim 0.1-0.3\,L^*$
\citep{ffp+05,hwf+05,cmi+06,sbk+06}.  It has also been proposed that
GRB\,050911 occurred in a galaxy cluster at $z=0.1646$ \citep{bsm+06},
that the old IPN burst GRB\,790613 was associated with an elliptical
galaxy at $z=0.09$ \citep{gno+05}, and that GRB\,060502b was ejected
from a bright galaxy at $z=0.287$ \citep{bpc+06}.  These low redshifts
have been used to argue for long progenitor lifetimes, $\gtrsim 4$
Gyr, and against a substantial population of short GRBs at high
redshift (e.g., \citealt{gp06,ngf06,hgw+06}).  They also set the
energy scale of short GRBs at $\sim 10^{48}-10^{49}$ erg (e.g.,
\citealt{sbk+06}).

Despite these initial results, there is tentative evidence that some
short GRBs may occur at higher redshifts.  This includes the proposed
association of GRB\,050813 with a galaxy cluster at $z\sim 1.8$
\citep{ber06}, a photometric redshift for GRB\,060121 of $z\sim 1.7$
or $\sim 4.6$ based on the afterglow optical/near-IR spectral energy
distribution \citep{ltf+06,ucg+06}, and limits on galaxy brightness of
$\gtrsim 19$ mag in error boxes of some poorly-localized short bursts
(typical size of $20$ arcmin$^{2}$; \citealt{sch06}).  Determining
with greater confidence whether a high redshift population in fact
exists, and how it relates to the low redshift short GRBs, remains an
open issue, with implications for the burst energetics, progenitor
lifetimes, and rate estimates.

Here we present optical observations of nine well-localized ($<6''$
radius) short GRBs discovered in the past year, and find that eight
are likely associated with faint galaxies, $R\sim 23-26.5$ mag (the
remaining host has $R\approx 21$ mag).  We show by comparison to the
previously-detected hosts (with $R\sim 17-22$ mag and $z\lesssim
0.5$), as well as the hosts of long GRBs and large field galaxy
samples, that these new host galaxies likely reside at $z\sim 1$.
Indeed, we present spectroscopic redshifts for the four {\it
brightest} hosts of $z\approx 0.4-1.1$.  These observations establish
for the first time that at about $1/3$ of all short GRBs originate at
high redshift, and that some bursts produce $10^{50}-10^{52}$ erg in
their prompt emission, at least two orders of magnitude larger than
the low redshift short bursts.  Most importantly, with this new high
redshift sample, we provide tighter constraints on the progenitor age
distribution than previously possible, and find that viable models
include a wide lognormal distribution with $\tau_*\sim 4-8$ Gyr, or
power law distributions, $P(\tau)\propto\tau^n$, with $-1\lesssim
n\lesssim 0$.

\section{Observations}
\label{sec:obs}

The prompt emission properties and X-ray afterglow positions of the
seven bursts discussed in this paper are provided in
Table~\ref{tab:grbs}.  The table also includes the properties of the
four previous short bursts with measured redshifts.  We consider here
only events for which X-ray or optical afterglow positions are
available, providing positional uncertainties better than $4.5''$
radius, and therefore a low probability of chance associations.  We
note that the prompt and X-ray afterglow properties of some of these
bursts are discussed in detail in the published literature.
GRB\,051210: \citet{lmf+06}; GRB\,060121: \citet{ucg+06},
\citet{dls+06}, and \citet{ltf+06}; GRB\,060313: \citet{rvp+06}; and
GRB\,060502b: \citet{bpc+06}.

Before addressing the individual bursts we note that in the context of
the popular model of NS-NS or NS-BH binaries the progenitors may
experience a kick, leading to mergers outside of the host galaxies.
The range of offsets depends on the distributions of kick velocities,
merger times, and host masses, but reasonable values are $\sim 10-100$
kpc \citep{fwh99,bpb+06}.  This translates to an angular distance of
about $6-60''$ at $z\approx 0.1$, or about $1.5-15''$ at $z\gtrsim
0.7$.  While kicks may provide an obstacle to secure associations when
sub-arcsecond positions are not available, we stress that in the
existing sample of short GRBs with secure associations the offsets are
small -- GRB\,050709: 3.8 kpc \citep{ffp+05}, GRB\,050724: 2.6 kpc
\citep{bpc+05}, and GRB\,051221a: 0.8 kpc \citep{sbk+06}.  In
addition, as we show below, when precise optical positions are
available for the new sample, the bursts invariably coincide at high
confidence level with faint galaxies.  If these bursts were ejected
from nearby galaxies there is no reason why they should always land on
an unrelated galaxy.  This, and the fact that not all progenitors are
expected to experience a significant kick in the first place,
indicates that kicks do not provide a significant source of
contamination.  Below we provide an assessment of the brightest
galaxies near each object and their associated probability of chance
coincidence, compared to the faint galaxies coincident with the
optical/X-ray afterglow positions.

Reduction of the Gemini data discussed below was performed using the
{\tt gemini} package in IRAF (for bias subtraction, flat-fielding, and
frame co-addition).  Magellan optical and near-IR observations were
reduced using standard IRAF routines, including for the latter dark
frame subtraction and fringe correction.  Throughout the paper we use
the standard cosmological parameters $H_0=70$ km s$^{-1}$ Mpc$^{-1}$,
$\Omega_m=0.27$, and $\Omega_\Lambda=0.73$.

\subsection{GRB\,051210}
\label{sec:051210}

Optical observations of this burst were obtained with the Low
Dispersion Survey Spectrograph (LDSS3) on the Magellan/Clay 6.5-m
telescope starting 19.4 hr after the burst for a total of 1200 s in
$r$-band.  These observations revealed a faint, extended object within
the \swift/XRT error circle at RA=\ra{22}{00}{40.93},
Dec=\dec{-57}{36}{47.1} (J2000; \citealt{gcn4330}).

We obtained a deeper observation of this burst with the LDSS3
instrument on 2006 Jan.~05 UT for a total exposure time of 1950 s in
$r$-band.  We detect the same extended object and measure its
brightness to be $r_{\rm AB}=24.04\pm 0.15$ mag in comparison to the
SDSS standard stars Feige 22 and G\,162-66; see
Figure~\ref{fig:hosts}.  No other sources are detected in the error
circle to a $3\sigma$ limit of $r_{\rm AB}>24.9$ mag.  We further note
that the nearest galaxies which are brighter than this putative host
(with 21.6 and 20.4 mag) are located $23''$ and $39''$ from the center
of the error circle, respectively (or, 115 and 195 kpc at $z\sim
0.3-0.5$).  The expected number of such objects at these offsets is
about 2 and 1.5, respectively \citep{bsk+06}.  Thus, the large offsets
and the order unity probability of chance coincidence suggest that
they are not likely to be associated with the burst.

We further undertook spectroscopic
observations\footnotemark\footnotetext{All Gemini observations in this
paper were obtained as part of programs GN-2005B-Q-6, GN-2006A-Q-14,
GN-2006B-Q-21, GS-2006A-Q-8, and GS-2006B-Q-12.} of the putative host
with the Gemini Multi-Object Spectrograph (GMOS; \citealt{hja+04})
mounted on the Gemini-South 8-m telescope on four consecutive nights
beginning on 2006 Dec.~20.04 UT.  A total of 9600 s were obtained
using the nod-and-shuffle mode with the R400 grating at central
wavelengths of 7250 and 7550 \AA.  The data were reduced using the
{\tt gemini} package in IRAF, while rectification and sky subtraction
were performed using the method and software described in
\citet{kel03}.  Wavelength calibration was performed using CuAr arc
lamps and air-to-vacuum and heliocentric corrections were applied.
The final combined spectrum covers $5000-9500$ \AA\ at a resolution of
about 7 \AA.  We detect weak continuum emission in each individual
spectrum, but no clear emission or absorption features are detected in
the combined spectrum.  The lack of detectable [\ion{O}{2}]$\lambda
3727$ emission (if the host is star forming) indicates $z\gtrsim
1.55$, while the lack of a clear Balmer/4000\AA\ break (if the host is
early-type) indicates $z\gtrsim 1.4$; we use the latter as a robust
lower limit on the redshift.

\subsection{GRB\,051227}

The optical afterglow was initially found in a pair of observations
obtained 10.4 and 12.5 hr after the burst with the VLT
\citep{gcn4407}.  We contemporaneously observed the burst position 
with GMOS on the Gemini-North 8-m telescope starting 13.9 hr after the
burst for a total exposure time of 1500 s in $r$-band, and confirmed
the presence of the optical source with $r_{\rm AB}=25.00\pm 0.12$
mag.  The position of the source is RA=\ra{08}{20}{58.11},
Dec=\dec{+31}{55}{32.0} (J2000) with an uncertainty of $0.08''$ in
each coordinate relative to SDSS.

Additional observations with GMOS were obtained 38.6 and 62.4 hr after
the burst for total exposures of 1500 and 1800 s, respectively, and
confirmed that the object has faded between the first and second
observations.  The brightness of the object remains constant between
the second and third observation, indicating the presence of the host
galaxy.  From the final observation we measure for the host $r_{\rm
AB}=25.78\pm 0.15$ mag (Figure~\ref{fig:hosts}).  The positional
offset between the afterglow and host is only $0.05\pm 0.02''$.  The
probability of chance coincidence at such a radius and brightness
level is about $2\times 10^{-4}$.  We note that a brighter galaxy
($r_{\rm AB} =22.28\pm 0.05$ mag) is located $4.6''$ away from the
optical afterglow position, but its probability of chance coincidence
is about $20\%$ \citep{bsk+06}, significantly larger than for the
underlying galaxy.  Either way, the redshift of this galaxy is
$z=0.714$ \citep{gcn4409}, higher than for the previous short bursts.

\subsection{GRB\,060121}
\label{sec:060121}

The optical afterglow was found by several groups starting 2 hr after
the burst.  Details are provided in \citet{ucg+06} and \citet{ltf+06}.
These authors find that the afterglow has an unusually red $R-K$
color, suggestive of extinction and/or a Lyman break.  The preferred
redshift is $z\sim 1.7$ or $\sim 4.6$ \citep{ucg+06,ltf+06}.  In
addition, {\it Hubble Space Telescope} (HST) observations with the
Advanced Camera for Surveys (ACS) and the Near Infra-red Camera and
Multi-Object Spectrometer (NICMOS) revealed an extended object
coincident with the position of the afterglow with ${\rm
F606W_{AB}}=27.0\pm 0.3$ mag and ${\rm F160W_{AB}}=24.5\pm 0.2$ mag
\citep{ltf+06}.

We observed the position of the afterglow with GMOS on Gemini-North
starting on 2006 Feb.~1.54 UT for a total exposure time of 1080 s in
$r$-band and 1050 s in $i$-band.  We detect the galaxy noted by
\citet{ltf+06}, and measure its brightness at $r_{\rm AB}=26.2\pm 0.3$
mag and $i_{\rm AB}>25.9$ mag ($3\sigma$) relative to several nearby
stars with SDSS photometry (\citealt{ceh+06}; see
Figure~\ref{fig:hosts}).

In addition, we obtained the ACS and NICMOS data from the HST archive
and processed the images using the {\tt multidrizzle} routine
\citep{fh02} in the {\tt stsdas} package of IRAF.  The NICMOS images
were first re-processed with an improved dark frame created from the
HUDF using the IRAF task {\tt calnica} in the {\tt nicmos} package.
We measure for the host galaxy ${\rm F606W_{AB}}=27.2\pm 0.3$ mag and
${\rm F160W_{AB}}=24.8\pm 0.1$ mag, consistent with the values given
by \citet{ltf+06}.  Images of the host are shown in
Figure~\ref{fig:060121}.

We note that our detection of the host in $r$-band ($\lambda_{\rm
eff}=630$ nm) is about 1 mag brighter than the F606W flux
($\lambda_{\rm eff}=590$ nm).  This is possibly indicative of a Lyman
break at $\lambda\approx 610$ nm, or $z\approx 4$, in good agreement
with the redshift estimated from the afterglow colors.  At this
redshift, the isotropic-equivalent energy is $E_{\gamma,{\rm
iso}}\approx 1.5\times 10^{53}$ erg, substantially larger than that of
any other short GRB to date (see also \citealt{ucg+06}).

Finally, for the nearby red galaxies noted by \citet{ltf+06} (see
Figure~\ref{fig:060121}) we measure from our Gemini data and the
NICMOS data, $(i-H)_{\rm AB}>2.7$, $>3.0$, $>2.1$, and $>2.3$ mag.
From the ACS data we find a $3\sigma$ limit of $V>27.6$ mag in a
$0.6''$ aperture, leading to colors of $(V-H)_{\rm AB}>4.4$, $>4.6$,
$>3.8$, and $>4.0$ mag.  The nearest of these red galaxies is located
$8.7''$ away from the afterglow position, or about 70 kpc at $z\sim
1$.  These galaxies represent an over-density by about a factor of 20
\citep{ltf+06}, but the large separation likely indicates that they
are not related to the burst.  We note that even if they are related,
they likely reside at $z>1$ \citep{ltf+06}.

\subsection{GRB\,060313}

The optical afterglow was discovered with the VLT, the Danish 1.5-m
telescope, the \swift\ UV/optical telescope (UVOT), and our Gemini
observations \citep{gcn4871,gcn4874,gcn4876,gcn4877}.  We observed the
position of the burst with GMOS on Gemini-South starting 71 min after
the burst for a total of 1800 s in $r$-band.  We clearly detect the
afterglow with $r_{\rm AB}=19.99\pm 0.02$ mag relative to several
nearby USNO stars (the systematic uncertainty is $0.18$ mag).
Follow-up observations with GMOS were obtained 1.01 d (900 s exposure)
and 2.02 d (1500 s exposure) after the burst, confirming that the
source has faded to $22.47\pm 0.07$ mag and $23.58\pm 0.14$ mag,
respectively; see Figure~\ref{fig:060313}.  Finally, we observed the
afterglow position on 2006 Mar.~22 UT (10 d after the burst) for a
total exposure time of 1800 s, but did not detect the afterglow to a
$3\sigma$ limit of $r_{\rm AB}>24.7$ mag (Figure~\ref{fig:060313}).  A
faint galaxy is detected about $0.4''$ from the position of the
afterglow in HST/ACS observations with ${\rm F775W_{AB}}=26.2\pm 0.2$
mag and ${\rm F475W_{AB}}=26.8\pm 0.2$ mag (Fox et al. in prep).  The
probability of chance coincidence is only $4\times 10^{-3}$
\citep{bsk+06}.

The nearest bright galaxy has $r_{\rm AB}\approx 18.7 $ mag and is
located about $27''$ away (or about 80 kpc at $z\sim 0.2$) from the
optical afterglow position.  The probability of chance coincidence for
this galaxy is about 0.06, significantly larger than for the faint
galaxy.  Moreover, the detection of a bright optical afterglow from
this burst requires a circumburst density, $n\gtrsim 10^{-3}$
cm$^{-3}$ (e.g., \citealt{sbk+06}), which is unlikely at such a large
offset from the host galaxy, where we would expect densities similar
to the intergalactic medium\footnotemark\footnotetext{If the burst
occurred in a globular cluster associated with the nearby galaxy the
density may be sufficient to produce an afterglow.}.  We therefore
consider this galaxy to be a chance association.

Finally, we obtained from the European Southern Observatory archive
VLT observations taken with the Infrared Spectrometer And Array Camera
(ISAAC) on 2006 Mar.~21.41 ($K_s$-band; 1320 s), Mar.~29.99 ($J$-band;
800 s) and Mar.~30.10 UT ($H$-band; 650 s).  No object is detected at
the position of the afterglow to $3\sigma$ limits of $K_{\rm AB}>22.9$
mag, $H_{\rm AB}>21.0$ mag, and $J_{\rm AB}>20.9$ mag relative to a
nearby 2MASS star.

\subsection{GRB\,060502b}

Initial optical observations revealed a single object within the XRT
error circle \citep{gcn5066,gcn5071}.  We obtained spectroscopy of
this object with GMOS on Gemini-North and showed that it is an M giant
star \citep{gcn5071}.  We further imaged the position of the burst
with GMOS starting 16.8 hr after the burst for a total exposure time
of 1500 s in $r$-band.  In addition to the star noted above we detect
a faint object within the XRT error circle with $r_{\rm AB}=23.95\pm
0.13$ mag relative to USNO-B (with a systematic uncertainty of 0.35
mag).  HST/ACS observations reveal that this object has a stellar
point spread function and is hence unlikely to be the host (Fox et
al. in prep.)

\begin{figure}
\epsscale{1}
\includegraphics[angle=0,width=7.3in]{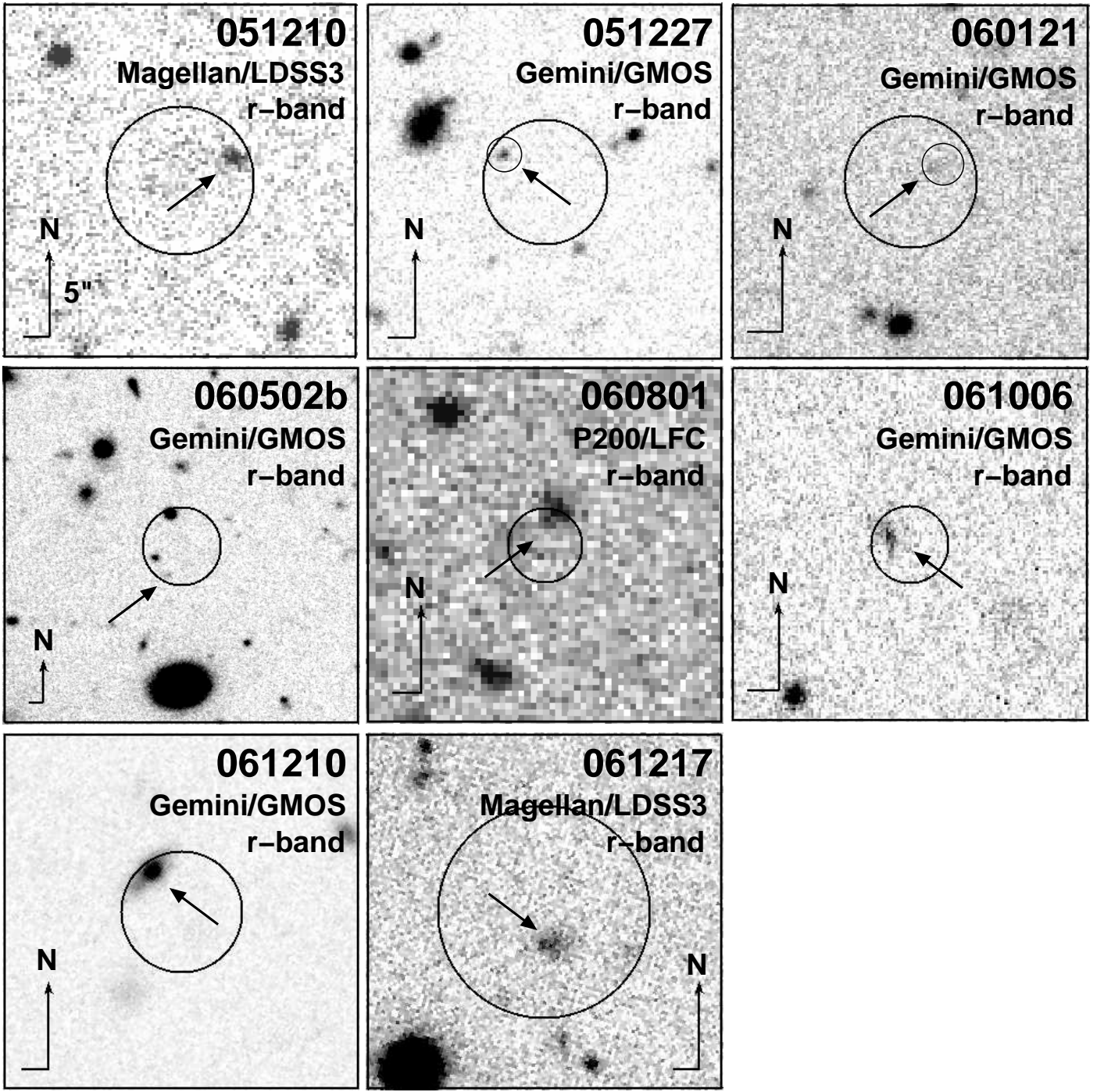}
\caption{Ground-based images from Magellan and Gemini of several short
GRB hosts.  All images are $20''$ on a side, with the exception of
GRB\,060502b which is twice as large.  The large circles mark the XRT
error regions, while smaller circles mark the positions of the optical
afterglows (when available).  Arrows mark the positions of the hosts.
For GRB\,060502b, the bright galaxy to the south of the XRT position
has been proposed as the host \citep{bpc+06}, but we note that there
is a faint galaxy within the XRT error circle (see also
\citealt{bpc+06}).
\label{fig:hosts}}
\end{figure}

\begin{figure}
\epsscale{1}
\includegraphics[angle=0,width=7.3in]{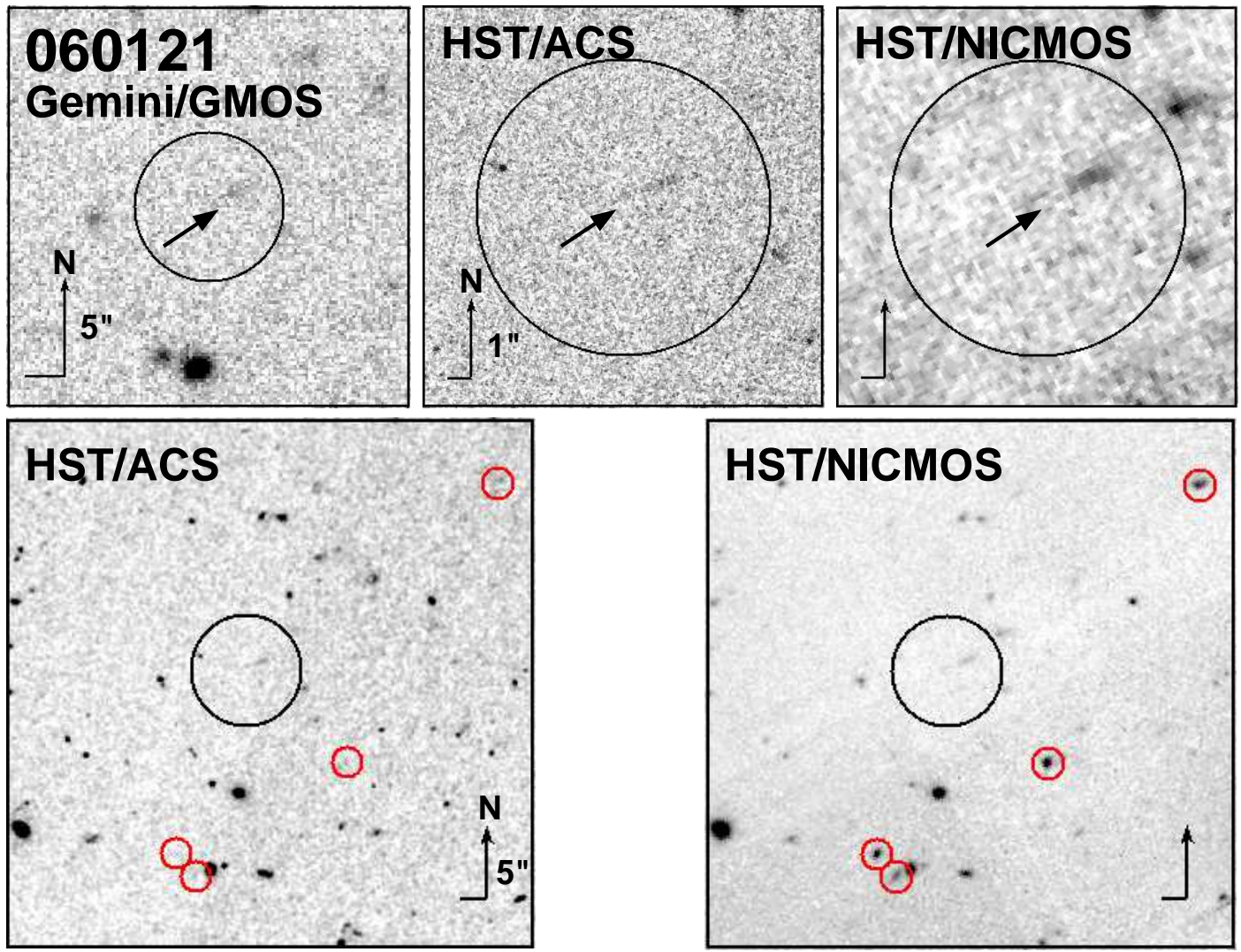}
\caption{Gemini $r$-band and {\it Hubble Space Telescope} ACS/F606W
and NICMOS/F160W observations of the host galaxy of GRB\,060121.  The
black circles mark the XRT error region.  The two bottom panels
provide a larger view of the field with four nearby very red galaxies
marked by red circles.  See also \citet{ltf+06}.
\label{fig:060121}}
\end{figure}

\begin{figure}
\epsscale{1}
\includegraphics[angle=0,width=7.3in]{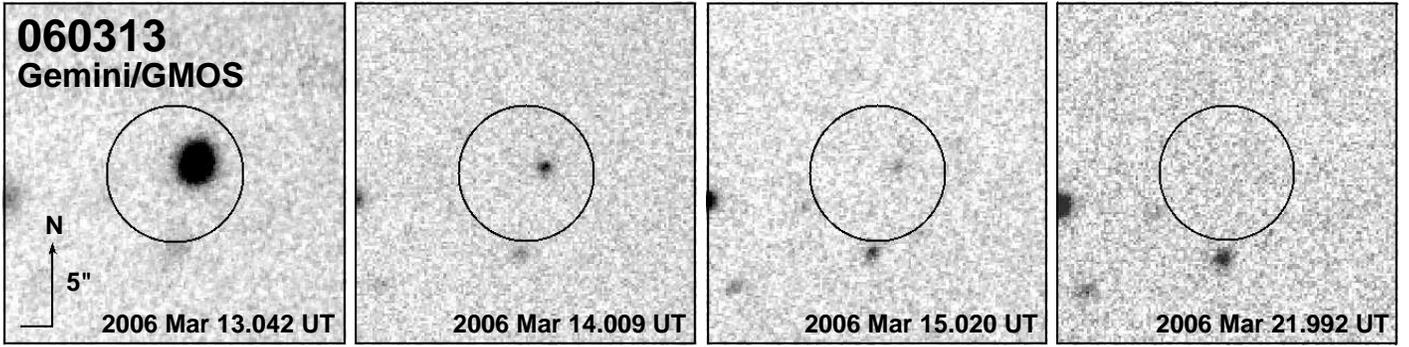}
\caption{Gemini $r$-band observations of the afterglow of GRB\,060313.
The fading behavior is evident.  No host galaxy is detected at the
position of the afterglow 10 days after the burst, to a $3\sigma$
limit of $r>24.7$ mag.  However, a faint galaxy is detected within
$0.4''$ of the afterglow position in HST/ACS images with ${\rm
F775W(AB)} =26.2\pm 0.2$ mag (Fox et al. in prep).  We note that the
source ellipticity in the initial epoch is due to the image quality.
\label{fig:060313}}
\end{figure}

A second $r$-band observation with GMOS was obtained 40.8 hr after the
burst for a total exposure time of 1500 s.  Due to improved seeing
conditions ($0.5''$ vs.~$0.95''$ in the first image) we detect an
additional faint source, not clearly visible in our first epoch, for
which we measure $r_{\rm AB}=25.22\pm 0.18$ mag (with a systematic
uncertainty of 0.4 mag); see Figure~\ref{fig:hosts}.  This object was
also noted by \citet{bpc+06}, who proposed instead that the host is a
bright galaxy $17.5''$ (or about 70 kpc) south of the XRT error circle
(Figure~\ref{fig:hosts}).  In this scenario the large offset requires
a progenitor kick of $v>55$ km s$^{-1}$ \citep{bpc+06}.  This proposed
association is based on a probability of chance coincidence of about
$0.03$ (see \S\ref{sec:spec}).

\subsection{GRB\,060801}
\label{sec:060801}

Optical observations revealed four objects within the initial XRT
error circle ranging in brightness from $R\approx 22$ to 24.6 mag, of
which none revealed any variability between 0.5 and 1.5 d after the
burst \citep{gcn5384,gcn5386,gcn5392}.  Only two of these sources are
located within the revised XRT error circle with $R=23.7$ mag (source
"B") and 24.6 mag (source "D") and extended morphologies
\citep{gcn5392}.  We obtained imaging observations of the burst with
the Large Format Camera on the Hale 200-inch telescope starting 16.0
hr after the burst for a total exposure time of 1500 s in $r$-band.
Photometry of the two extended sources relative to several nearby SDSS
stars indicates $r_{\rm AB}=23.20\pm 0.11$ mag and $24.1\pm 0.3$ mag,
respectively, somewhat brighter than the magnitude quoted in the GCN
circular.  We note that the nearest galaxies with significantly
brighter magnitudes ($r_{\rm AB}\approx 19.8-20.5$ mag) are located
$40-70''$ away from the XRT position.  The probability of chance
coincidence for these galaxies is of order unity.

\begin{figure}
\epsscale{1}
\includegraphics[angle=0,width=7.3in]{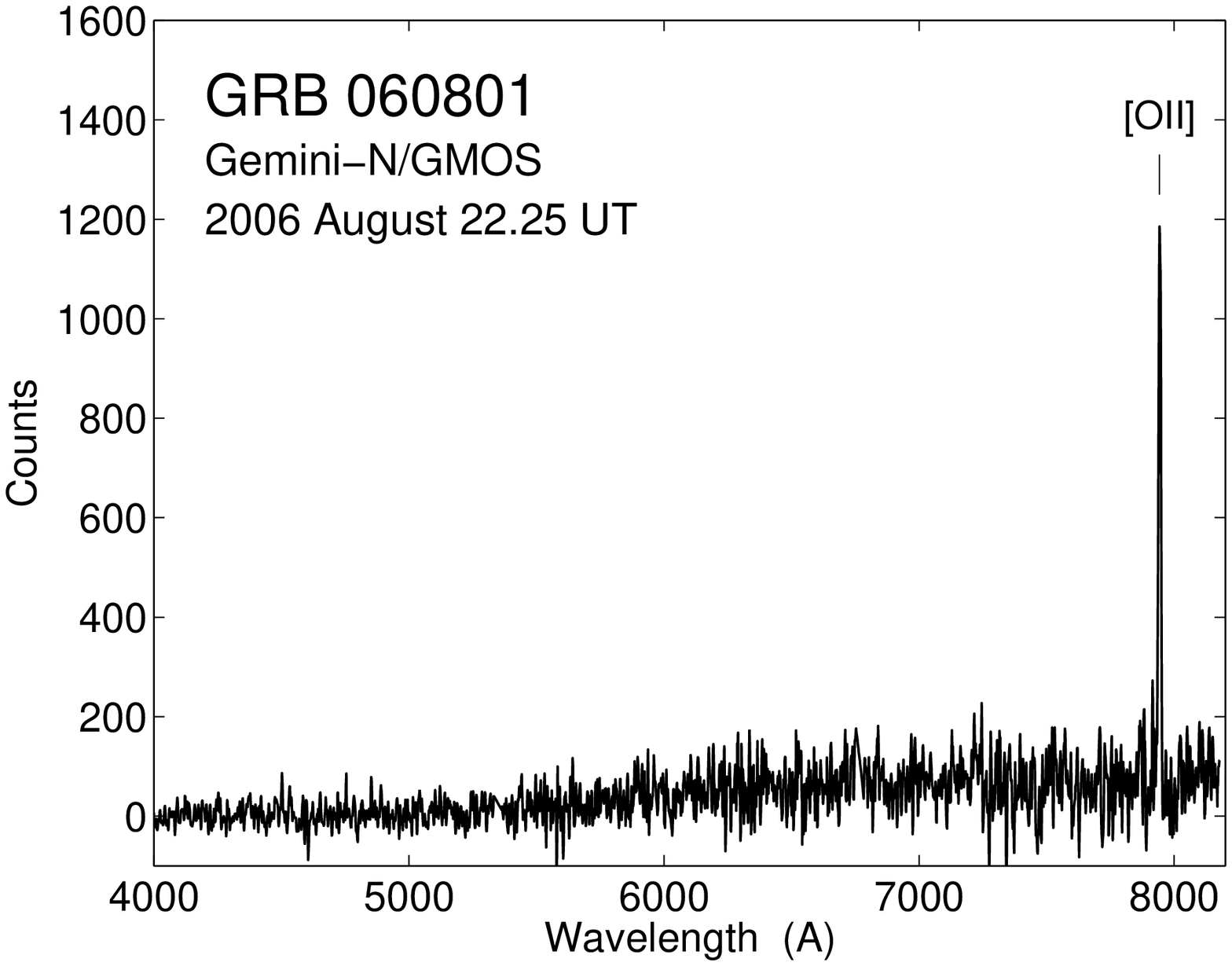}
\caption{Gemini/GMOS spectrum of the putative host galaxy of
GRB\,060801, smoothed with a 3-pixel boxcar.  We detect a single
bright emission line, which we identify as the [\ion{O}{2}]$\lambda
3727$ doublet at $z=1.1304$.  
\label{fig:060801}}
\end{figure}

We obtained spectroscopic observations of source B with GMOS on
Gemini-North on 2006 Aug.~22.25 UT, for a total exposure time of 1800
s with the R400 grating at a central wavelength of 6050 \AA.  The data
were reduced using the {\tt gemini} package in IRAF, while
rectification and sky subtraction were performed using the method and
software described in \citet{kel03}.  Wavelength calibration was
performed using CuAr arc lamps and air-to-vacuum and heliocentric
corrections were applied.  The spectrum covers $4000-8200$ \AA\ at a
resolution of about 7 \AA.  We detect weak continuum emission and a
single broad (${\rm FWHM}\approx 11$ \AA) emission line at a
wavelength of 7943.19 \AA, which we identify as the barely-resolved
[\ion{O}{2}]$\lambda 3727$ doublet at $z=1.1304$
(Figure~\ref{fig:060801}).  We note that all things being equal,
source D, which is a factor of two fainter, is likely to reside at an
even higher redshift.

At $z=1.1304$, the putative host galaxy has an absolute magnitude,
$M_B\approx -21$ mag, or $L_B\approx L^*$ compared to the luminosity
function of $z\sim 1.1$ galaxies in the DEEP2 survey \citep{wfk+06}.
In addition, the isotropic equivalent energy of the burst at this
redshift is $E_{\gamma,{\rm iso}}=(2.7\pm 0.3)\times 10^{50}$ erg.

\subsection{GRB\,061006}

Optical observations with the VLT revealed a single object within the
XRT error circle of this burst, which faded by about 0.5 mag between
14.6 and 38.4 hr after the burst \citep{gcn5705,gcn5718}.  We observed
the position of the afterglow with GMOS on Gemini-South starting 3.6 d
after the burst.  A total of 900 s and 1440 s were obtained in $i$-
and $r$-band, respectively.  We detect a faint, extended object
coincident with the afterglow at RA=\ra{07}{24}{07.75},
Dec=\dec{-79}{11}{55.3} (J2000) with an uncertainty of about $0.16''$
relative to several nearby 2MASS stars (Figure~\ref{fig:hosts}).
Photometry of this object relative to USNO-B indicates $r_{\rm AB}
=24.18\pm 0.09$ mag (with a systematic uncertainty of 0.26 mag) and
$i_{\rm AB}=23.11\pm 0.09$ mag (with a systematic uncertainty of 0.30
mag).  We note that there are no other significantly brighter galaxies
within about $45''$ of the afterglow position.

In addition, we observed the position of this source with Persson's
Auxiliary Nasmyth Infrared Camera (PANIC) on the Magellan/Baade 6.5-m
telescope on 2006 Oct.~8.39 UT in $Y$-band for a total of 960 s.  We
detect an extended source coincident with the optical position with
$Y=22.0\pm 0.2$ mag using a $Y$-band zero-point of 25.06 mag for PANIC
measured on 2006 Aug.~21 (C.~Burns private communication).  We
estimate the uncertainty in the zero-point to be about $30\%$.

\begin{figure}
\epsscale{1}
\includegraphics[angle=0,width=7.3in]{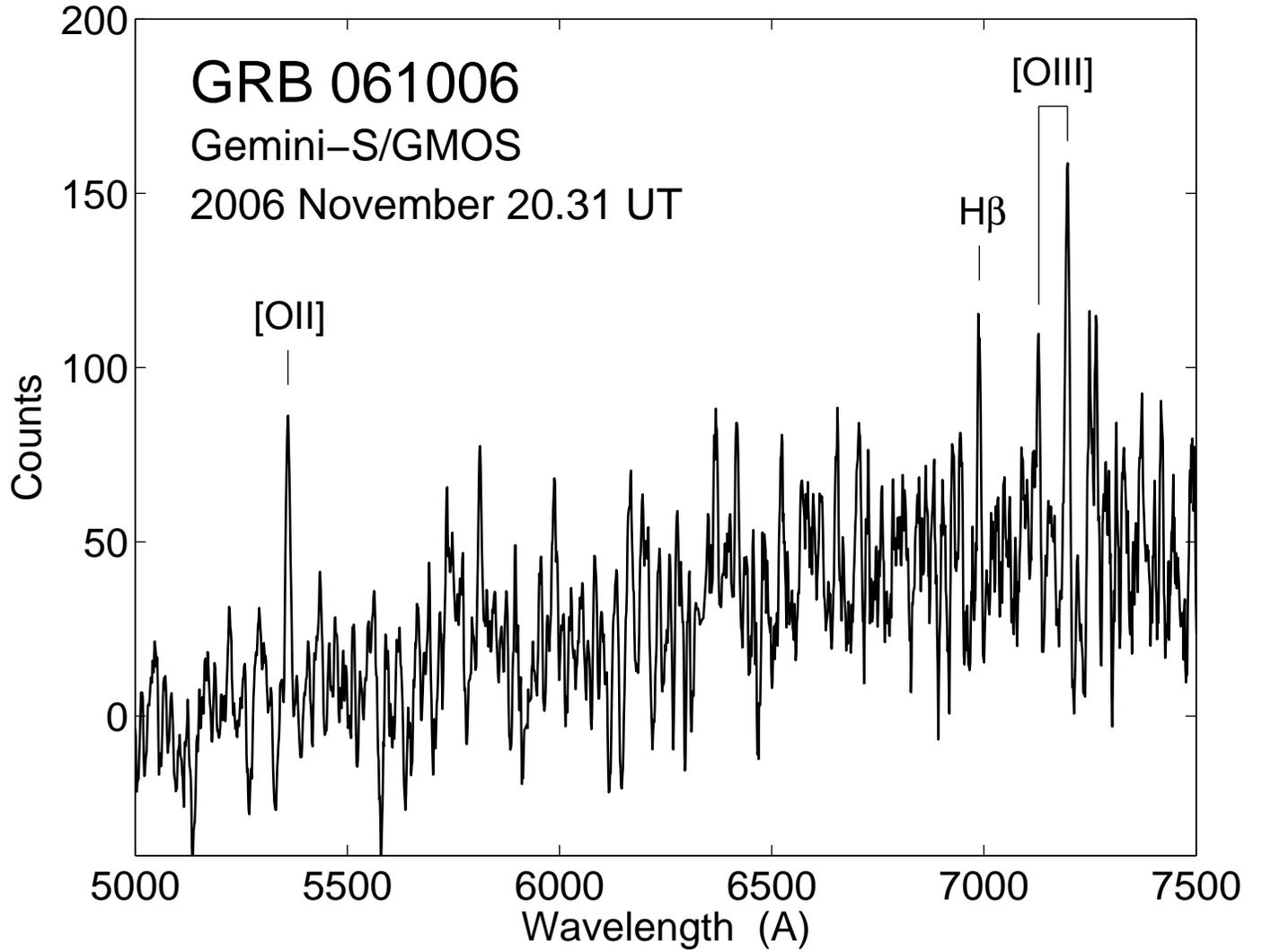}
\caption{Gemini/GMOS spectrum of the putative host galaxy of
GRB\,061006, smoothed with a 7-pixel boxcar.  We detect several
emission lines corresponding to [\ion{O}{2}]$\lambda 3727$, H$\beta$,
[\ion{O}{3}]$\lambda 4959$, and [\ion{O}{3}]$\lambda 5007$ at
$z=0.4377\pm 0.0002$.
\label{fig:061006}}
\end{figure}

Finally, we obtained spectroscopic observations of the putative host
galaxy with GMOS on Gemini-South on 2006 Nov.~20.31 UT for a total
exposure time of 3600 s using the nod-and-shuffle mode with the R400
grating at a central wavelength of 7200 \AA.  The data were reduced
using the {\tt gemini} package in IRAF, while rectification and sky
subtraction were performed using the method and software described in
\citet{kel03}.  Wavelength calibration was performed using CuAr arc
lamps and air-to-vacuum and heliocentric corrections were applied.
The spectrum covers $4900-9200$ \AA\ at a resolution of about 7 \AA.
We detect weak continuum emission and several emission lines
corresponding to [\ion{O}{2}]$\lambda 3727$, H$\beta$,
[\ion{O}{3}]$\lambda 4959$, and [\ion{O}{3}]$\lambda 5007$ at
$z=0.4377\pm 0.0002$ (Figure~\ref{fig:061006}).

At this redshift the putative host galaxy has an absolute magnitude,
$M_B\approx -18.6$ mag, or $L_B\approx 0.1L^*$ compared to the
luminosity function of $z\sim 0.5$ galaxies in the DEEP2 survey
\citep{wfk+06}.  In addition, the isotropic equivalent energy of the
burst at this redshift is $E_{\gamma,{\rm iso}}=(6.9\pm 0.5)\times
10^{50}$ erg.

\subsection{GRB\,061210}

We observed the BAT error circle with GMOS on Gemini-North on two
separate occassions, 2.1 and 25.6 hr after the burst.  Digital image
subtraction using the ISIS software package \citep{al98} revealed no
variable sources to a limit of $r_{\rm AB}>23.5$ mag at the time of
the first observation.  The subsequent detection of two X-ray sources
within the BAT error circle \citep{gcn5921} allowed us to propose
candidate host galaxies, of which the brightest has $r_{\rm
AB}=21.00\pm 0.02$ mag and is located at RA=\ra{09}{38}{05.36},
Dec=\dec{+15}{37}{18.8} (J2000) with an uncertainty of about $0.25''$
relative to USNO-B (Figure~\ref{fig:hosts}).  The XRT source
containing this galaxy eventually faded \citep{gcn5983}, confirming
that it is likely the host galaxy of GRB\,061210.  We note that the
nearest galaxies which are brighter than the putative host (by about
1.8 mag) are located about $45''$ and $60''$ away from the X-ray
afterglow position.

\begin{figure}
\epsscale{1}
\includegraphics[angle=0,width=7.3in]{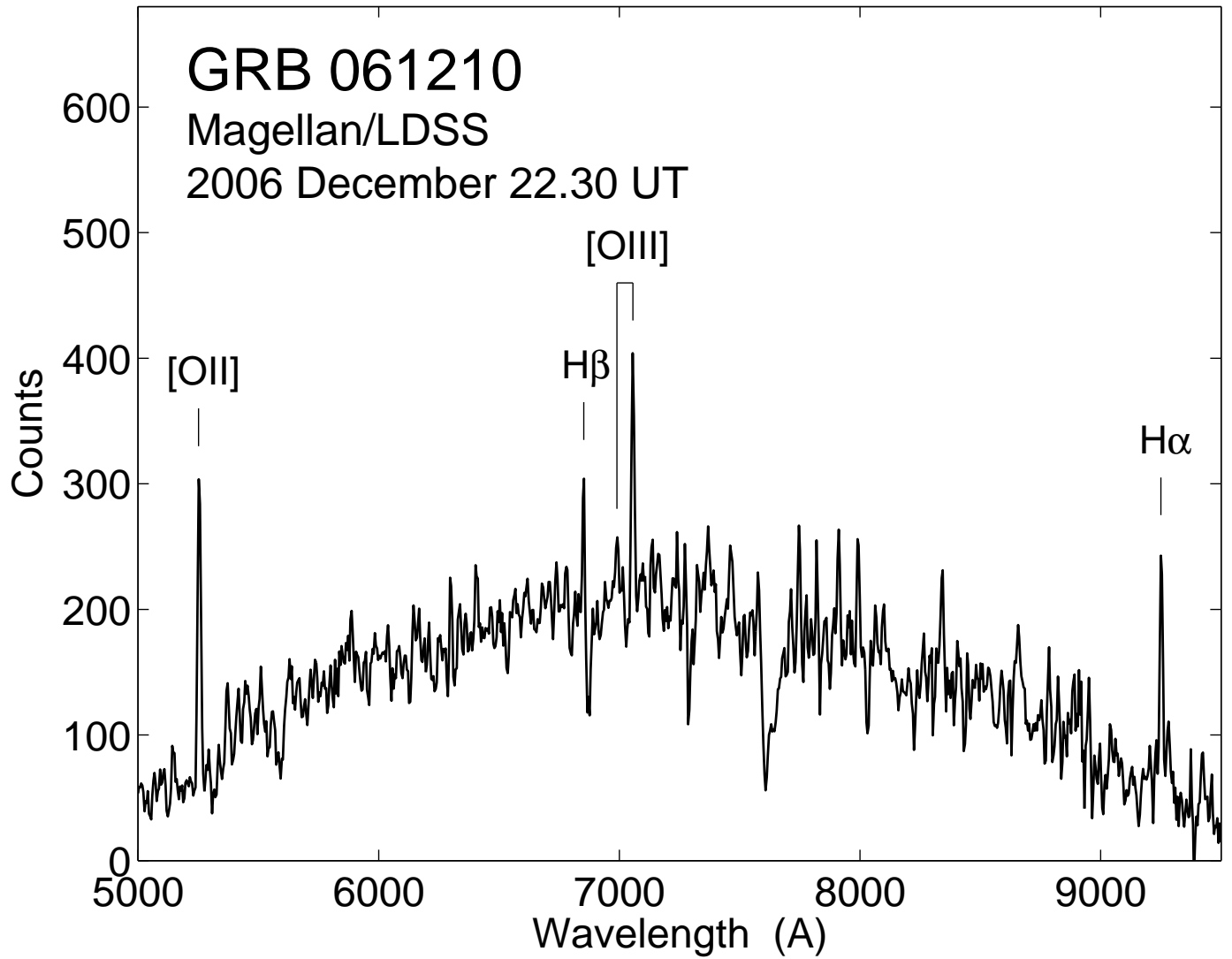}
\caption{Magellan/LDSS3 spectrum of the putative host galaxy of
GRB\,061210, smoothed with a 3-pixel boxcar.  We detect several
emission lines corresponding to [\ion{O}{2}]$\lambda 3727$, H$\beta$,
[\ion{O}{3}]$\lambda 4959$, [\ion{O}{3}]$\lambda 5007$, and H$\alpha$
at $z=0.4095\pm 0.0001$.
\label{fig:061210}}
\end{figure}

We obtained spectroscopic observations of the putative host with LDSS3
on the Magellan/Clay 6.5-m telescope on 2006 Dec.~22.30 UT, for a
total exposure time of 4400 s.  The data were reduced using standard
IRAF packages, while rectification and sky subtraction were performed
using the method and software described in \citet{kel03}.  Wavelength
calibration was performed using HeNeAr arc lamps and air-to-vacuum and
heliocentric corrections were applied.  The spectrum covers
$3500-9800$ \AA\ at a resolution of about 7 \AA.  We detect several
emission lines corresponding to [\ion{O}{2}]$\lambda 3727$, H$\beta$,
[\ion{O}{3}]$\lambda 4959$, [\ion{O}{3}]$\lambda 5007$, and H$\alpha$
at $z=0.4095\pm 0.0001$ (Figure~\ref{fig:061210}).

At this redshift the putative host galaxy has an absolute magnitude,
$M_B\approx -20.4$ mag, or $L_B\approx 1.5 L^*$ compared to the
luminosity function of $z\sim 0.4$ galaxies in the DEEP2 survey
\citep{wfk+06}.  In addition, the isotropic equivalent energy of the
burst at this redshift is $E_{\gamma,{\rm iso}}=(4.6\pm 0.8)\times
10^{50}$ erg.

\subsection{GRB\,061217}

We observed the XRT error circle of GRB\,061217 \citep{gcn5947} with
LDSS3 on the Magellan 6.5-m telescope on 2006 Dec.~21.36 UT for a
total of 600 s in $r$-band.  This led to the identification of an
extended object with $r_{\rm AB}=23.33\pm 0.07$ located at
RA=\ra{10}{41}{39.08}, Dec=\dec{-21}{07}{28.7} (J2000) with an
uncertainty of about $0.28''$ relative to USNO-B
(Figure~\ref{fig:hosts}).  No brighter galaxies are detected within a
radius of about $80''$.

\begin{figure}
\epsscale{1}
\includegraphics[angle=0,width=7.3in]{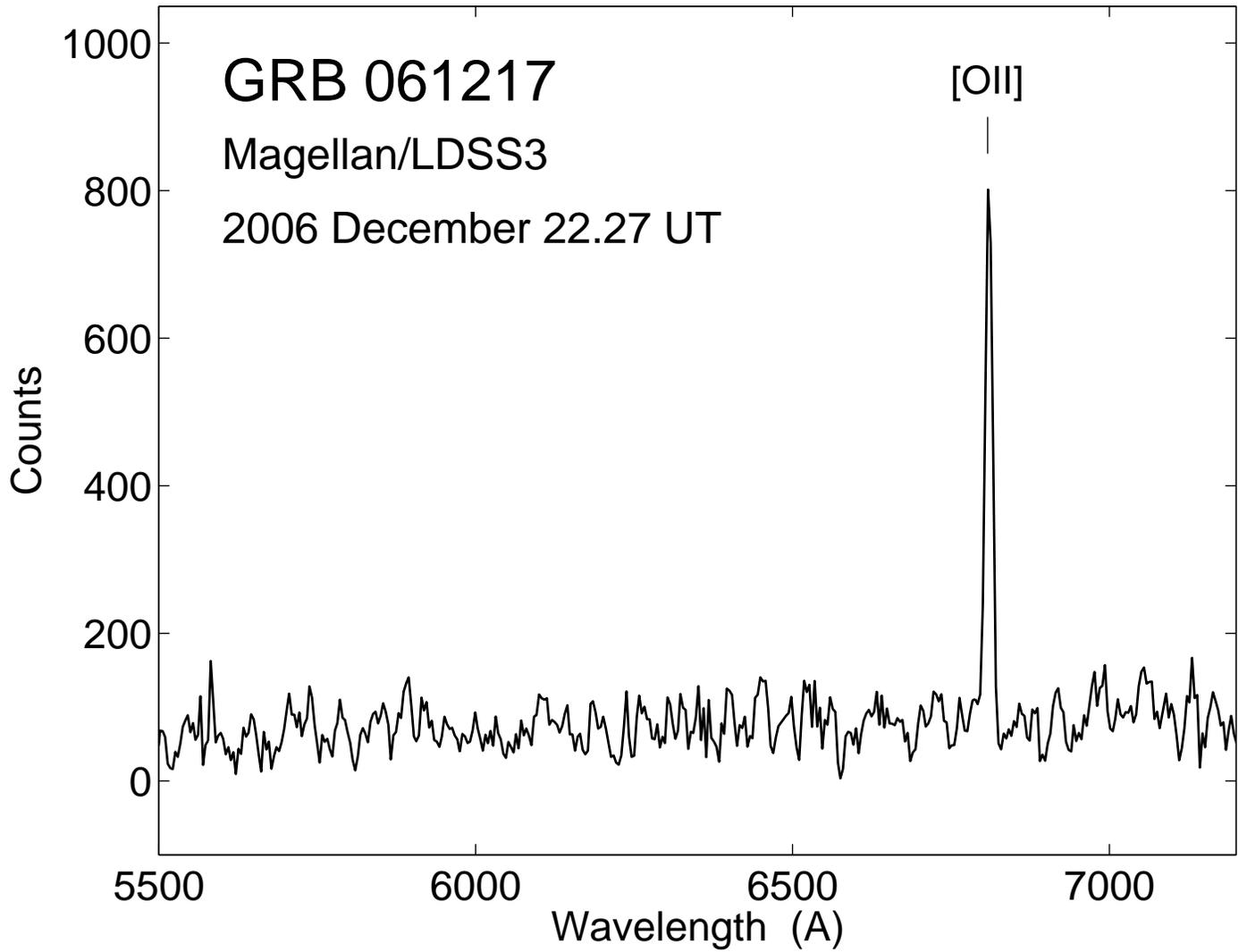}
\caption{Magellan/LDSS3 spectrum of the putative host galaxy of
GRB\,061217, smoothed with a 3-pixel boxcar.  We detect a single
bright emission line, which we identify as the [\ion{O}{2}]$\lambda
3727$ doublet at $z=0.8270$.  
\label{fig:061217}}
\end{figure}

We obtained spectroscopic observations of the putative host with LDSS3
on 2006 Dec.~22.24 UT, for a total exposure time of 4100 s.  The data
were reduced using standard IRAF packages, while rectification and sky
subtraction were performed using the method and software described in
\citet{kel03}.  Wavelength calibration was performed using HeNeAr arc
lamps and air-to-vacuum and heliocentric corrections were applied.
The spectrum covers $4500-9500$ \AA\ at a resolution of about 6 \AA.
We detect a single bright and resolved emission line which we identify
as the [\ion{O}{2}]$\lambda 3727$ doublet at $z=0.8270$
(Figure~\ref{fig:061217}).

At this redshift the putative host galaxy has an absolute magnitude,
$M_B\approx -19.6$ mag, or $L_B\approx 0.5 L^*$ compared to the
luminosity function of $z\sim 0.8$ galaxies in the DEEP2 survey
\citep{wfk+06}.  In addition, the isotropic equivalent energy of the
burst at this redshift is $E_{\gamma,{\rm iso}}=(8.3\pm 1.4)\times
10^{49}$ erg.

\section{A $z\sim 1$ Host Galaxy Population}
\label{sec:spec}

In order to address the redshift distribution of the new short GRBs in
a robust way we consider in addition to the full sample of nine events
the subset of unambiguous short bursts: 051210, 060313, 060502b,
060801, 061006, 061210, and 061217.  The remaining two bursts (051227
and 060121) are most likely in the short duration category
\citep{gcn4401,dls+06} but have formal $T_{90}$ durations of $\gtrsim
2$ s.  The discussion of host redshift likelihoods below applies to
all events, but for the purpose of burst statistics we consider both
samples separately where appropriate.

\begin{figure}
\epsscale{1}
\includegraphics[angle=0,width=7.3in]{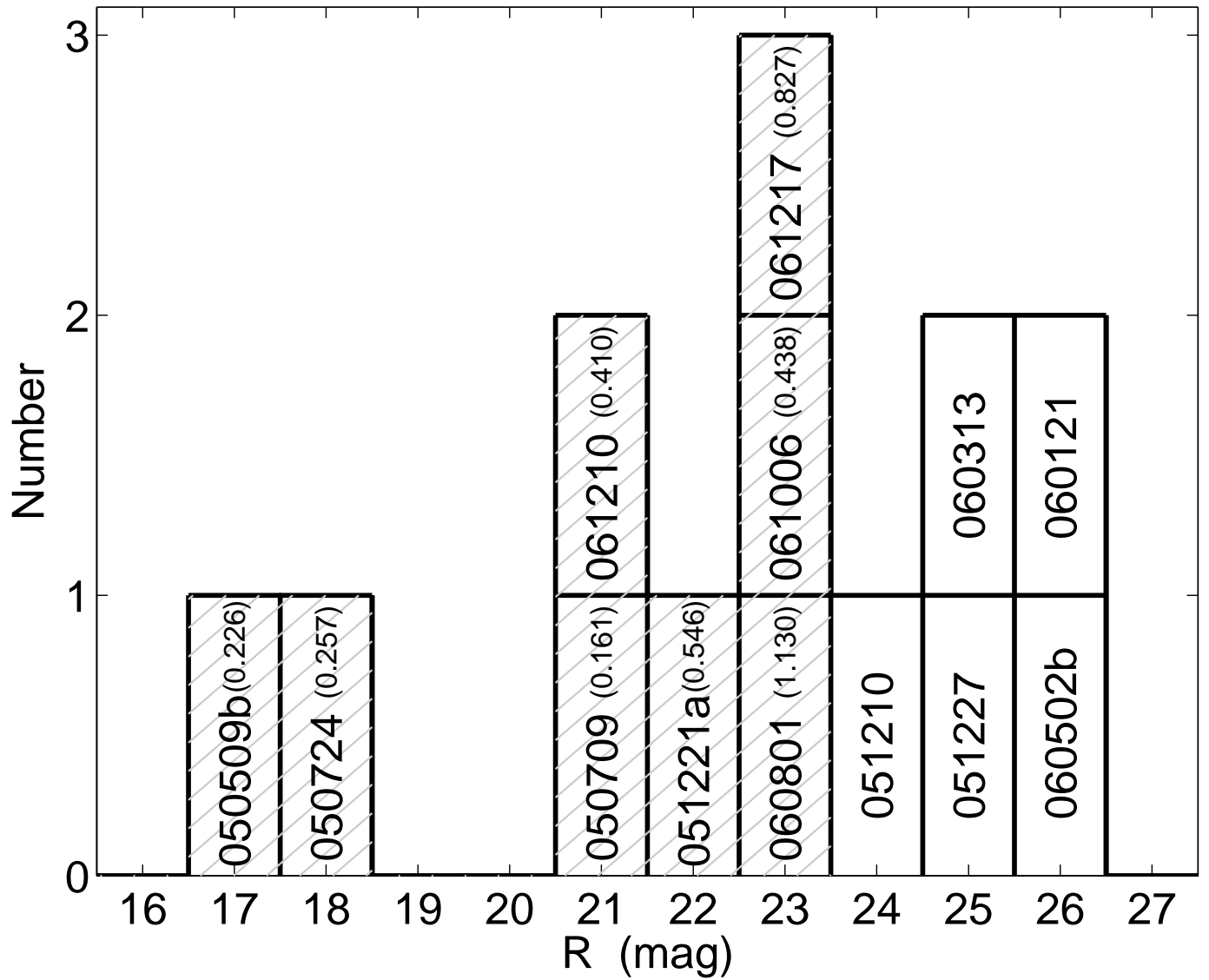}
\caption{Histogram of observed $R$-band magnitudes for the host
galaxies of short GRBs.  Hatched bars mark the hosts for which we have
a measured redshift.  Clearly, the low-redshift hosts are at the
bright end of the distribution, and their redshifts are therefore not
representative of the entire sample.  On the other hand, the only host
with $z>1$ appears to be more representative of the faint hosts,
suggesting these galaxies are also located at $z\gtrsim 1$.
\label{fig:mags}}
\end{figure}

The observed magnitudes of the candidate host galaxies (with the
excpetion of GRB\,061210), corrected for Galactic extinction, range
from $R=22.6$ to $26.3$ mag; see Table~\ref{tab:grbs}.  The
distribution of magnitudes for our sample, as well as the low redshift
hosts detected previously is shown in Figure~\ref{fig:mags}.  Overall,
the host magnitudes range up to $R\approx 17$ mag.  The two brightest
hosts (050509b and 050724) are elliptical galaxies and would stand out
from the distribution even more if we considered their near-IR
brightness.  We find that the median host brightness is $\langle
R\rangle=23.0\pm 0.8$ mag, nearly two magnitudes brighter than the
median value of $\langle R\rangle=24.8\pm 0.5$ mag for the hosts of
long GRBs.

\begin{figure}
\epsscale{1}
\includegraphics[angle=0,width=7in]{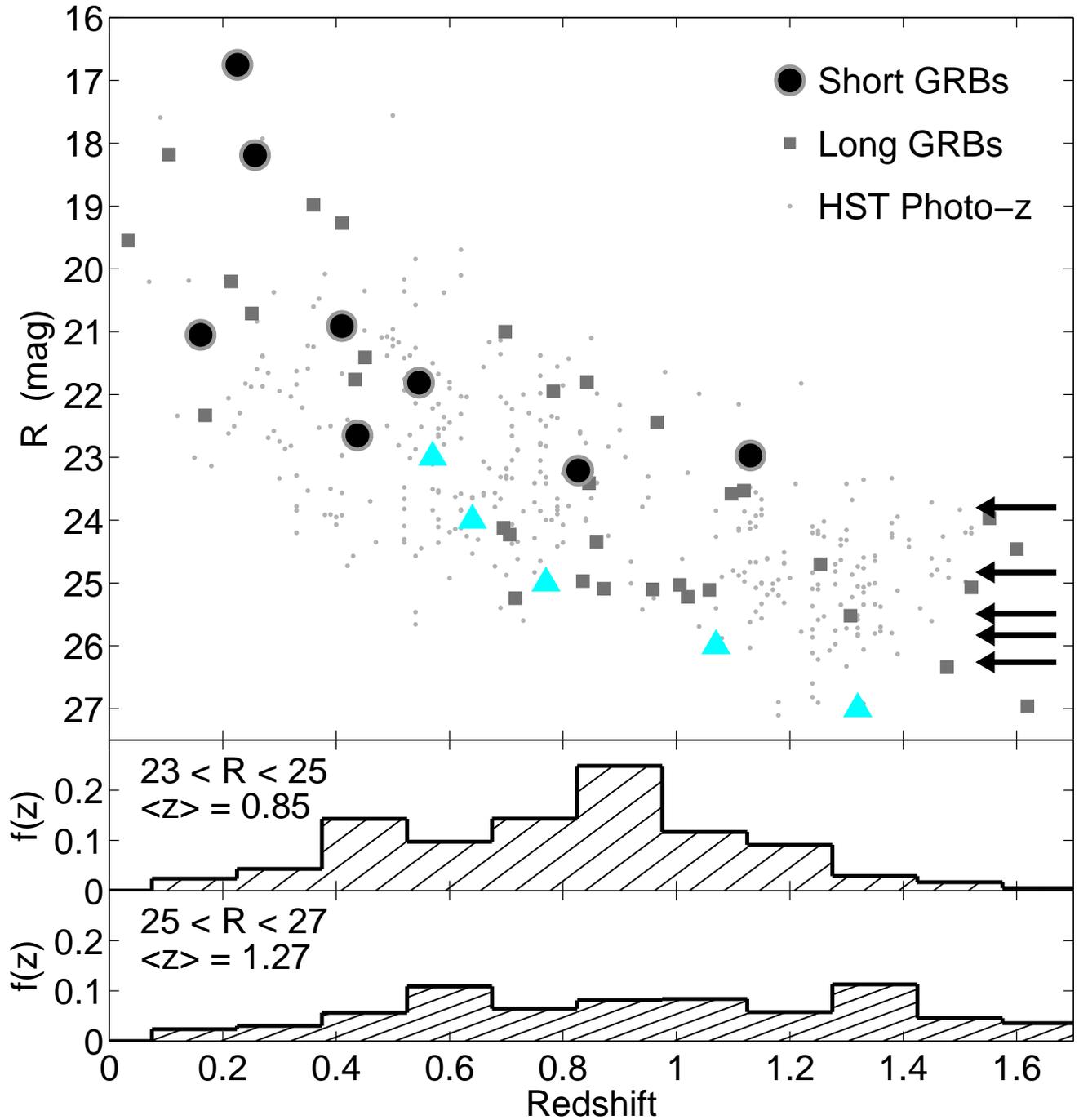}
\caption{Host galaxy $R$ magnitudes (corrected for Galactic
extinction; \citealt{sfd98}) plotted versus redshift for short GRBs
(solid black circles, arrows), long GRBs (gray squares), and galaxies
in the HST/ACS Early Release Observation fields VV 29 (UGC 10214) and
NGC 4676 \citep{bfb+06}.  In all samples we observe a trend of
increasing apparent magnitude with redshift.  The upward pointing
triangles indicate the median redshift of a galaxy sample complete to
the appropriate magnitude limit \citep{cbs+06}.  For $R\gtrsim 26$ mag
appropriate for our sample the median redshift is about 1.1.  The
bottom panels show the redshift distributions of galaxies in two
magnitude bins from spectroscopic ($23<R<25$ mag;
\citealt{cbh+04,wwa+04}) and photometric ($25<R<27$ mag;
\citealt{cbs+06}) redshift surveys.  The clear magnitude-redshift 
relation for short GRB hosts suggests that the faint host galaxies
discussed in this paper are located at $z\sim 1$.  
\label{fig:zr}}
\end{figure}

The most crucial point demonstrated in Figure~\ref{fig:mags} is that
the four secure redshifts previously available ($z=0.1606-0.5465$)
belong to the four brightest host galaxies.  This is not surprising
given the relative ease of spectroscopic follow-up for galaxies with
$R\lesssim 22$ mag.  Similarly, the four redshifts presented in this
paper, $z\approx 0.4-1.1$, belong to the next four brightest hosts,
and the highest ones ($z=0.827$ and $z=1.130$) are measured for the
hosts with $R\approx 23$ mag.  Extending this trend to the rest of the
faint host sample, we conclude\footnotemark\footnotetext{We note that
for GRB\,060313 the limit on the redshift is $z\lesssim 1.7$ based on
the detection of the afterglow in the UVOT/UVW2 filter with
$\lambda_{\rm eff}\approx 2000$ \AA\ \citep{rvp+06}.  For GRB\,051210
the likely redshift is $z\gtrsim 1.4$ (\S\ref{sec:051210}).} that they
most likely reside at $z\sim 1$ and beyond; see Figure~\ref{fig:zr}.
In addition, we note that the new hosts with measured spectroscopic
redshifts continue the trend that short GRBs occur in $\sim L^*$
galaxies (Table~\ref{tab:grbs}).

If on the other hand we were to argue that the faint hosts are located
at a low redshift, $z\lesssim 0.5$, then the implied absolute
magnitudes would be $M_B\gtrsim -17$ mag, or $L\lesssim 0.01\,L^*$.
This is $10-100$ times fainter than the previously-detected low
redshift hosts, again pointing to a difference in the two host
populations.  We consider this possibility highly unlikely for two
primary reasons.  First, the two faintest hosts for which we do have a
spectroscopic redshift are located at $z=0.827$ and $z=1.130$ in $L^*$
galaxies; the remaining five hosts are even fainter and are therefore
likely to be at even higher redshifts (Table~\ref{tab:grbs}).
Similarly, for GRB\,060121 the probability that it is located at
$z<0.5$ is only $5\times 10^{-3}$ based on the afterglow SED
\citep{ucg+06}, and GRB\,051210 resides at $z\gtrsim 1.4$.  Second, 
even in the sample of long GRBs, which {\it are} thought to be biased
in favor of low-luminosity galaxies \citep{fls+06,sgb+06}, all
galaxies with $R>23$ mag are located at $z>0.7$ (Figure~\ref{fig:zr}).

Our conclusion that the hosts are located at $z\sim 1$ is further
supported by a comparison to large galaxy samples with spectroscopic
and photometric redshifts \citep{cbh+04,wwa+04,cbs+06}.  As shown in
Figure~\ref{fig:zr}, there is a clear trend of decreasing brightness
with redshift, which is also seen in the sample of short GRBs with a
measured redshift.  As mentioned above, even the subset of long GRB
hosts with $R>23$ mag reside exclusively at $z>0.7$.  Similarly, the
median redshift of galaxies with $23<R<25$ mag in the GOODS region is
about 0.85 \citep{cbh+04,wwa+04}, while galaxies with $25<R<27$ mag in
the HUDF have a median (photometric) redshift of about 1.3
\citep{cbs+06}.  The HUDF sample also indicates that for a limiting
magnitude of $R\approx 26$ mag, appropriate for our sample of faint
short GRB hosts, the expected median redshift of a typical galaxy
sample is about 1.1 \citep{cbs+06}.  Making the reasonable assumption
that the hosts without a measured redshift are drawn the from the
general population of galaxies, based on the various arguments 
provided above, we conclude that they likely reside at $z\gtrsim 1$.

To provide formal constraints we note that in the GOODS and HUDF
galaxy samples $70\%$ of the galaxies with $23<R<27$ mag are located
at $z>0.7$.  For our sample of seven unambiguous short bursts, there
is therefore a $97\%$ probability that at least three reside at
$z>0.7$ (as confirmed by the measured spectroscopic redshifts).
Considering the full sample of nine events, the corresponding number
is four galaxies.  For a limiting redshift of $z=0.55$, corresponding
to the highest redshift previously measured for a short GRB, the
fraction of field galaxies with $23<R<27$ mag above this redshift is
$84\%$.  Therefore, there is a $97\%$ probability that four bursts in
the unambiguous sample reside at $z>0.55$, or five out of the full
sample.  Based on the secure host identifications, it appears that
short GRB hosts are drawn from the bright end of the galaxy luminosity
function (Figure~\ref{fig:zr}).  This implies that the inferred
fraction at $z>0.7$, or alternatively the inferred median redshift,
are probably even higher than the values above.

Finally, we stress that four of the nine bursts have optical afterglow
positions, making the host associations highly likely.  In fact,
taking a conservative positional uncertainty of $0.2''$ radius for
these bursts, the probability of chance coincidence at $R=23$ mag (26
mag) is only about $3.5\times 10^{-4}$ ($3.5\times 10^{-3}$) using the
cumulative galaxy number counts in the HUDF and GOODS
\citep{bsk+06}.  The median brightness of these four hosts is $\langle
R\rangle\approx 25.2$ mag, indicating that our inference of a high
redshift origin is robust.  As noted above, with such low
probabilities of chance coincidence the possibility that these bursts
were instead ejected from nearby brighter galaxies is highly unlikely.

For the other five events, two of the putative hosts (051210 and
061210) are the only source within the XRT error circle to the limit
of our observations, while for GRBs 060502b, 060801, and 061217 there
may be more than one galaxy within the error circle, but they are all
{\it fainter} than our putative hosts (Figure~\ref{fig:hosts} and
\citealt{bpc+06}).  While the chance coincidence probability is high
($\sim 0.1-1$) for galaxies of similar or brighter magnitude within
the XRT error circles, the lack of viable nearby bright alternatives
for GRBs 051210, 060108, 061210, and 061217 (\S\ref{sec:obs})
indicates that our identified hosts are secure (or that the hosts are
even fainter).

This leaves the host of GRB\,060502b, which was proposed to be a
bright galaxy $17.5''$ (or about 70 kpc) south of the XRT error circle
(Figure~\ref{fig:hosts}; \citealt{bpc+06}).  In this scenario the
large offset requires a progenitor kick of $v>55$ km s$^{-1}$
\citep{bpc+06}.  This proposed association is based on both a
probability of chance coincidence of about $3\%$ for the putative
bright host (compared to order unity for the faint galaxies {\it
within} the XRT error circle), as well as its similarity to the hosts
of GRBs 050509b and 050724.  However, based on the sample presented
here we suggest that the ejection scenario is not required.  First,
GRBs 050724, 050709, and 051221a did not have a substantial offsets
from their hosts.  Moreover, we show here that the bursts with precise
optical afterglow positions are coincident with faint hosts.  Thus, in
none of the secure cases do we find evidence for a required offset.
In fact, there is a $33\%$ chance probability that one out of the
thirteen bursts with positional accuracies better than $\sim 6''$,
will be located within $17.5''$ of a bright galaxy as the one proposed
by \citet{bpc+06}, given a single-trial probability of $3\%$.  With
such a high probability of chance coincidence it cannot be
convincingly argued that GRB\,060502b was ejected from its host.

Second, as can be seen in Figure~\ref{fig:mags} the hosts of GRBs
050509b and 050724 do not appear to be representative of the general
short GRB host population.  We therefore argue that the coincidence of
at least half of the short GRBs with galaxies fainter than $R\approx
23$ mag suggests that the host of GRB\,060502b is most likely one of
the faint galaxies within the XRT error circle.  While there is a
higher probability of chance coincidence for such faint galaxies, this
association removes the need for a (model-dependent) progenitor kick.
With the existence of the new sample of faint hosts, future evidence
for significant offsets (and hence progenitor kicks) will have to rely
on a large statistical sample, rather than individual cases, or a
direct determination of the burst redshift from an absorption
spectrum, which coincides with the redshift of an offset galaxy.

\section{Discussion}
\label{sec:disc}

We present optical observations from Magellan, Gemini, and HST for
nine short GRBs, of which four have sub-arcsecond positions from
optical afterglow detections, and the rest are localized to better
than $6''$ radius based on the X-ray afterglow.  We find that eight of
the nine bursts appear to be associated with galaxies fainter than
$R\approx 23$ mag (and the remaining with an $R=21$ mag galaxy), in
contrast to previous short GRBs that were associated with galaxies
brighter than $R\approx 22$ mag at $z\lesssim 0.5$.  This suggests
that the new hosts reside at higher redshifts, and indeed our
spectroscopic redshifts are in the range $z\approx 0.4-1.1$, with the
two faintest hosts residing at the highest redshifts.

Using the conservative subset of unambiguous short bursts we conclude
at the $97\%$ confidence level that at least three short GRBs from our
sample originated at $z>0.7$; for the full sample, the corresponding
number is four.  To this sample we can add GRB\,050813, which is most
likely associated with a galaxy cluster at $z\sim 1.8$, and is
certainly located beyond $z=0.72$ \citep{ber06,pbc+06}.  Thus, we
conclude that in either the full or the unambiguous samples at least
$1/3$ of the short bursts are located at $z>0.7$, with an upper bound
of about $60\%$.  This conclusion is therefore completely robust
against ambiguity about the nature of some of these short GRBs.  We
note that such a redshift distribution has been predicted in the
context of NS-NS and NS-BH progenitors by \citet{bpb+06}, with peak
redshifts of $z\sim 1$ and $\sim 1.5$, respectively.

We now address the implications of our results for the age and energy
distributions of short GRBs.  Analysis of the redshift and luminosity
distributions of the first few short GRBs led to the conclusion that
the progenitors experience a long time delay prior to the GRB
explosion \citep{gp06,ngf06,hgw+06}.  The favored models required a 
delay of $\gtrsim 4$ Gyr, or a power law distribution, $P(\tau)\propto
\tau^n$ with $n\gtrsim -1/2$ \citep{ngf06}.  In addition, \citet{zr06} 
find $n\gtrsim 3/2$ from the ratio of short GRBs in early- and 
late-type galaxies at low redshift.  The discovery of a high redshift
population now implies that not all progenitors are several Gyr old.
In the context of a single age distribution, and assuming a single 
power-law luminosity function, models with a lognormal age
distribution are required to be broad ($\sigma\gtrsim 1$) and with a
characteristic age, $\tau_*\sim 4-8$ Gyr (see for example Figure 3 of
\citealt{ngf06}).  Narrow lognormal distributions ($\sigma\sim 0.3$)
cannot reproduce both the low and high redshift samples.

Models with a power law distribution, on the other hand, are required
to have $-1\lesssim n\lesssim 0$.  The lower bound predicts about
$60\%$ of all short GRBs at $z>0.7$, consistent with our upper limit
of $\sim 60\%$, while the upper bound on $n$ predicts about $25\%$ at
$z>0.7$ consistent with our minimum estimate of $\sim 1/3$.  A
distribution with $n=3/2$ \citep{zr06} predicts less than $10\%$ of
the short GRBs at $z>0.7$, in conflict with our findings of $\sim
30-60\%$.  We note that the allowed range of $n$ values is still
consistent with existing data on the relative fraction of short GRBs
in cluster versus field early-type galaxies \citep{bsm+06,sb06}.
Additional spectroscopic redshifts are required for further refinement
of the age distribution, but the new limits already suggest that the
typical ages are shorter than previously deduced.

The existence of a population of short GRBs at $z\sim 1$ also
indicates that the energy release of some events may be larger than
previously suspected.  For the bursts with spectroscopic redshifts
presented in this paper, we find that the $\gamma$-ray fluences
(Table~\ref{tab:grbs}) correspond to isotropic equivalent energies of
$E_{\gamma,{\rm iso}}\sim 10^{50}-10^{51}$ erg.  For the remaining
bursts, assuming $z=1$, we find $E_{\gamma,{\rm iso}}\sim
10^{50}-10^{52}$ erg, and possibly $\sim 10^{53}$ erg for GRB\,060121
if it is located at $z\sim 4$ (\S\ref{sec:060121}).  This can be
contrasted with $E_{\gamma,{\rm iso}}\approx {\rm few}\times 10^{48}$
erg for GRBs 050509b, 050709, and 050724 \citep{bpc+05,ffp+05,bpp+06},
and a beaming-corrected $E_{\gamma}\approx 1.5\times 10^{49}$ erg for
GRB\,051221a \citep{bgc+06,sbk+06}.  If beaming corrections are not
significant, the inferred energies in excess of $\sim 10^{51}$ erg are
difficult to explain in models of $\nu\bar{\nu}$ annihilation
\citep{rr02}.  This may point to energy extraction via MHD processes
(e.g., \citealt{bz77}).  On the other hand, if the true energy release
of all short GRBs is about $10^{49}$ erg, then the required jet
opening angles for the high redshift bursts are about $10^\circ$.
This is consistent with the opening angle inferred for GRB\,051221a
\citep{bgc+06,sbk+06}.

Finally, we re-iterate that there is growing interest in the possible
detection of gravitational waves from short GRBs in the context of
compact object mergers \citep{dhh+06,ngf06}.  This is partly because
the detection of gravitational waves will provide insights about the
underlying system (e.g., degree of beaming, masses of the
constituents), while non-detection will rule out the binary merger
model.  Moreover, from the point of view of gravitational wave
detection of astrophysical sources, short GRBs provide a clean signal
thanks to directional and temporal information.  At the current
sensitivity of LIGO ($d\lesssim 20$ Mpc; \citealt{ct02}) it
is unlikely that short GRBs will be detected in the absence of
significant beaming and a low-luminosity population \citep{ngf06}.

However, the order of magnitude increase in sensitivity for Advanced
LIGO should broaden the science reach dramatically.  At the predicted
Advanced LIGO sensitivity, a binary with 1.4 M$_\odot$ constituents
would be detectable\footnotemark\footnotetext{With the two LIGO
observatories and the Virgo detector operational, the detection
requires a signal-to-noise ratio ${\rm SNR}>7$, or each individual
detector to have ${\rm SNR}>7/sqrt(3)=4$.  Our maximum distance is for
an average over all possible inclinations of the binary, and over all
possible sky positions.} out to $\sim 520$ Mpc (here we follow
\citealt{dhh+06}).  If short GRBs are beamed \citep{bgc+06,sbk+06} the
three-station detectability increases to $\sim 580$ Mpc.  In the
optimal case of a face-on binary directly overhead, the maximum
distance is $\sim 1.3$ Gpc ($z=0.26$).  We also note that the addition
of a fourth observatory (e.g., AIGO\footnotemark\footnotetext{\tt
http://www.gravity.uwa.edu.au/docs/aigo\_prospectus.pdf}) would 
increase these values to $600/675/1500$ Mpc, respectively.

With the full sample of events found previously and presented in this
paper, we find that at most $1/3$ of the short bursts are located
within $z\lesssim 0.25$ (compared to 5/8 found by \citealt{gno+05} and
\citealt{ngf06}).  With the detectability distances quoted above, this
leads to an expected event rate for Advanced LIGO of about 6
yr$^{-1}$, using the BATSE all-sky rate of 170 yr$^{-1}$.
Alternatively, we note that of the rough observed rate of three bursts
per year at $z<0.3$, one (050709) is within the range of detectablity
of an Advanced LIGO network.  With a constant comoving density this
indicates that $\sim 10\%$ of the $z<0.3$ bursts would be detectable,
or an extrapolated all-sky rate of $\sim 2$ yr$^{-1}$.  Thus, even
with no additional correction factors for beaming and low luminosity
events we find an expected Advanced LIGO rate of $\sim 2-6$ yr$^{-1}$,
indicating that simultaneous operations of this network and a
$\gamma$-ray satellite are of crucial importance (see also
\citealt{ngf06}).

The observations presented in this paper move us a step closer to an
unbiased view of short GRBs.  In the near term, additional
spectroscopic redshifts for the faint host population are essential.
In addition to confirming the distance scale of these galaxies, the
spectra will also provide an indication of the host types (early
vs.~late), the age of the dominant stellar population, star formation
rates, and possibly associations with galaxy clusters or groups
\citep{bsm+06,sb06}.  This information, along with morphological
classification from HST observations (Fox et al.~in prep.), will allow
us to refine the determination of progenitor ages
\citep{gno+05,sb06,zr06}, as well as the distribution of energy
release and the expected rate for future gravitational wave
experiments.

\acknowledgements 

We thank Scott Barthelmy and Louis Barbier for information on
GRB\,051227, and Chris Belczynski, Vicky Kalogera, and Brad Schaefer
for their various comments.  E.B.~is supported by NASA through Hubble
Fellowship grant HST-01171.01 awarded by the Space Telescope Science
Institute, which is operated by AURA, Inc.~for NASA under contract NAS
5-26555.  A.G.-Y.~is support by NASA through Hubble Fellowship grant
HST-HF-01158.01-A.  Based in part on observations obtained at the
Gemini Observatory, which is operated by the Association of
Universities for Research in Astronomy, Inc., under a cooperative
agreement with the NSF on behalf of the Gemini partnership: the
National Science Foundation (United States), the Particle Physics and
Astronomy Research Council (United Kingdom), the National Research
Council (Canada), CONICYT (Chile), the Australian Research Council
(Australia), CNPq (Brazil) and CONICET (Argentina)

\clearpage
\begin{deluxetable}{llllcccccccl}
\tablecolumns{12}
\tabcolsep0.045in\footnotesize
\tablewidth{0pc}
\tablecaption{Short GRB Properties
\label{tab:grbs}}
\tablehead {
\colhead {GRB}             &
\colhead {Date}            &
\colhead {$T_{90}$}        &
\colhead {$F_\gamma\,^a$}  &
\colhead {RA}              &
\colhead {Dec.}            &
\colhead {Uncert.}         &
\colhead {OA?}             &
\colhead {$z$}             &
\colhead {$R\,^b$}         &
\colhead {$L_B$}           &
\colhead {Refs.}           \\
\colhead {}                &
\colhead {(UT)}            &
\colhead {(s)}             &
\colhead {(erg cm$^{-2}$)} &
\colhead {(J2000)}         &
\colhead {(J2000)}         &
\colhead {($''$)}          &
\colhead {}                &
\colhead {}                &
\colhead {(mag)}           &
\colhead {($L^*$)}         &
\colhead {}          
}
\startdata
051210  & 2005 Dec.~10.240 & 1.27     & $(8.1\pm 1.4)\times 10^{-8}$     & \ra{22}{00}{41.3}  & \dec{-57}{36}{48.2} & 4.2 & N & $\gtrsim 1.4$ & $23.80\pm 0.15$ & \nod  & 1      \\
051227  & 2005 Dec.~27.755 & 8.0$^c$  & $(2.3\pm 0.3)\times 10^{-7}$     & \ra{08}{20}{57.92} & \dec{+31}{55}{30.4} & 3.5 & Y & \nod   & $25.49\pm 0.15$ & \nod  & 2--4   \\
060121  & 2006 Jan.~21.934 & 1.97     & $(4.7\pm 0.4)\times 10^{-6}$     & \ra{09}{09}{52.13} & \dec{+45}{39}{44.9} & 3.7 & Y & \nod   & $26.26\pm 0.30$ & \nod  & 5--6   \\
060313  & 2006 Mar.~13.008 & 0.70     & $(1.1\pm 0.1)\times 10^{-6}$     & \ra{04}{26}{28.50} & \dec{-10}{50}{40.2} & 4.0 & Y & \nod   & $24.83\pm 0.20$ & \nod  & 7      \\
060502b & 2006 May 2.726   & 0.09     & $(4.0\pm 0.5)\times 10^{-8}$     & \ra{18}{35}{45.74} & \dec{+52}{37}{52.5} & 4.4 & N & \nod   & $25.83\pm 0.05$ & \nod  & 8--10  \\
060801  & 2006 Aug.~1.511  & 0.50     & $(8.1\pm 1.0)\times 10^{-8}$     & \ra{14}{12}{01.31} & \dec{+16}{58}{54.0} & 2.1 & N & 1.1304 & $22.97\pm 0.11$ & $1$   & 11--13 \\
061006  & 2006 Oct.~6.699  & 0.42$^d$ & $(1.4\pm 0.1)\times 10^{-6}$     & \ra{07}{24}{07.33} & \dec{-79}{11}{55.8} & 2.2 & Y & 0.4377 & $22.65\pm 0.09$ & $0.1$ & 14--16 \\
061210  & 2006 Dec.~10.514 & 0.19$^e$ & $(1.1\pm 0.2)\times 10^{-6}$     & \ra{09}{38}{05.24} & \dec{+15}{37}{16.5} & 3.4 & N & 0.4095 & $21.00\pm 0.02$ & $1.5$ & 17--19 \\
061217  & 2006 Dec.~17.153 & 0.21     & $(4.6\pm 0.8)\times 10^{-8}$     & \ra{10}{41}{39.10} & \dec{-21}{07}{26.9} & 6.0 & N & 0.8270 & $23.33\pm 0.07$ & $0.5$ & 20--21 \\\hline\hline
050509b & 2005 May 9.167   & 0.04     & $(9.5\pm 2.5)\times 10^{-9}$     & \ra{12}{36}{13.58} & \dec{+28}{59}{01.3} & 9.3 & N & 0.226  & $16.75\pm 0.05$ & $5$   & 22--23 \\
050709  & 2005 Jul.~9.942  & 0.07     & $(2.9\pm 0.4)\times 10^{-7}\,^f$ & \ra{23}{01}{26.96} & \dec{-38}{58}{39.5} & 0.4 & Y & 0.1606 & $21.05\pm 0.07$ & $0.1$ & 24--26 \\
050724  & 2005 Jul.~24.524 & 3.0$^g$  & $(3.9\pm 1.0)\times 10^{-7}$     & \ra{16}{24}{44.36} & \dec{-27}{32}{27.5} & 0.5 & Y & 0.257  & $18.19\pm 0.03$ & $1$   & 27--29 \\
051221a & 2005 Dec.~21.077 & 1.40     & $(1.2\pm 0.1)\times 10^{-6}$     & \ra{21}{54}{48.62} & \dec{+16}{53}{27.2} & 0.2 & Y & 0.5465 & $21.81\pm 0.09$ & $0.3$ & 30--31 
\enddata
\tablecomments{Properties of the short GRB discussed in this paper,
including (i) GRB name, (ii) localization date, (iii) duration, (iv)
fluence, (v-vii) position of the X-ray afterglow including
uncertainty, (viii) whether an optical afterglow was detected, (ix) 
spectroscopic redshift, (x) host $R$-band magnitude corrected for
Galactic extinction \citep{sfd98}, (xi) rest-frame $B$-band 
luminosity, and (xi) relevant references.  The four bursts at the 
bottom of the table are the low redshift events previously 
identified. \\
$^a$ Fluence is in the $15-150$ keV energy band unless otherwise 
noted. \\
$^b$ Corrected for Galactic extinction \citep{sfd98}. \\
$^c$ The burst is likely in the short-duration category based both 
on the light curve similarity to GRB\,050724 with extended soft 
emission, and a negligible lag \citep{gcn4401}. \\
$^d$ This burst exhibits an extended soft tail with a duration 
of about 130 s \citep{gcn5704}; detection with Konus-Wind indicates 
a $20-2000$ keV fluence of $3.6\times 10^{-6}$ erg cm$^{-2}$ 
\citep{gcn5710}. \\
$^e$ This burst exhibits an extended soft tail with a duration 
of about 85 s \citep{gcn5905}. \\
$^f$ Fluence is in the $30-400$ keV band. \\
$^g$ The light curve is dominated by a 0.25 s hard spectrum spike, 
with a BATSE duration of $T_{90}=1.3$ s. \\
References: [1] \citet{lmf+06}; [2]
\citet{gcn4400}; [3] \citet{gcn4401}; [4] \citet{gcn4402}; [5]
\citet{gcn4550}; [6] \citet{gcn4565}; [7] \citet{rvp+06}; [8]
\citet{gcn5064}; [9] \citet{gcn5093}; [10] \citet{bpc+06}; [11]
\citet{gcn5381}; [12] \citet{gcn5381}; [13] \citet{gcn5389}; [14]
\citet{gcn5704}; [15] \citet{gcn5710}; [16] \citet{gcn5723}; [17]
\citet{gcn5905}; [18] \citet{gcn5921}; [19] \citet{gcn5983}; [20] 
\citet{gcn5930}; [21] \citet{gcn5947}; [22] 
\citet{gso+05}; [23] \citet{bpc+06}; [24] \citet{vlr+05}; [25] 
\citet{ffp+05}; [26] \citet{hwf+05}; [27] \citet{bcb+05}; [28] 
\citet{bpc+05}; [29] \citet{gbp+06}; [30] \citet{bgc+06}; [31] 
\citet{sbk+06}.}
\end{deluxetable}

\clearpage

\clearpage

\clearpage

\clearpage

\clearpage

\clearpage

\clearpage

\clearpage

\clearpage

\end{document}